\documentclass[copyright]{eptcs}
\providecommand{\event}{T. Neary, D. Woods, A.K. Seda and N. Murphy (Eds.):\\ The Complexity of Simple Programs 2008.}
\providecommand{\volume}{1}
\providecommand{\anno}{2009}
\providecommand{\firstpage}{149}
\usepackage{breakurl}

\usepackage{amsfonts}
\usepackage{amsmath}
\usepackage{amssymb}

\usepackage{amsfonts}
\usepackage{amsmath}
\usepackage{amssymb}
\usepackage{enumerate}
\usepackage{breakurl}

\newcommand{\PROOF}{\begin{proof}}

\newcommand{\QED}{\end{proof}}

\newcommand{\Dim}{{\mathrm{Dim}}}

\newcommand{\D}{{\mathcal{D}}}

\newcommand{\K}{{\mathrm{K}}}

\newcommand{\CH}{\mathcal{H}}

\numberwithin{equation}{section}

\newcommand{\I}{\mathcal{I}}
\newcommand{\dimfs}{\mathrm{dim}_\mathrm{FS}}
\newcommand{\Dimfs}{\mathrm{Dim}_\mathrm{FS}}

\title{A Divergence Formula for Randomness and Dimension\\
{\small (Short Version)}}
\author{Jack H. Lutz\footnote{This research was supported in part by National Science Foundation
 Grants 9988483, 0344187, 0652569, and 0728806 and by the Spanish
 Ministry of Education and Science (MEC) and the European Regional
 Development Fund (ERDF) under project TIN2005-08832-C03-02.}
\institute{Department of Computer Science, Iowa State University,
Ames, IA 50011 USA.
}
\email{lutz@cs.iastate.edu.
}
}
\def\titlerunning{A Divergence Formula for Randomness and Dimension}
\def\authorrunning{Jack H. Lutz}
\begin{document}
\date{}
\maketitle
\begin{abstract}

If $S$ is an infinite sequence over a finite alphabet $\Sigma$ and $\beta$
is a probability measure on $\Sigma$, then the {\it dimension} of $ S$ with
respect to $\beta$, written $\dim^\beta(S)$, is a constructive version of
Billingsley dimension that coincides with the (constructive Hausdorff)
dimension $\dim(S)$ when $\beta$ is the uniform probability measure.  This
paper shows that $\dim^\beta(S)$ and its dual $\Dim^\beta(S)$, the {\it strong
dimension} of $S$ with respect to $\beta$, can be used in conjunction with
randomness to measure the similarity of two probability measures
$\alpha$ and $\beta$ on $\Sigma$.  Specifically, we prove that the
{\it divergence formula}
\[
 \dim^\beta(R) = \Dim^\beta(R) =\frac{\CH(\alpha)}{\CH(\alpha) + \D(\alpha || \beta)}
\]
holds whenever $\alpha$ and $\beta$ are computable, positive probability
measures on $\Sigma$ and $R \in \Sigma^\infty$ is random with respect to
$\alpha$.  In this formula, $\CH(\alpha)$ is the Shannon entropy of $\alpha$,
and $\D(\alpha||\beta)$ is the Kullback-Leibler divergence between $\alpha$
and $\beta$.
\end{abstract}

\section{Introduction}\label{se:1}
The constructive dimension $\dim(S)$ and the constructive strong
dimension $\Dim(S)$ of an infinite sequence $S$ over a finite alphabet
$\Sigma$ are constructive versions of the two most important classical
fractal dimensions, namely, Hausdorff dimension \cite{Haus19} and packing
dimension \cite{Tricot82,Sull84}, respectively.  These two constructive
dimensions, which were introduced in \cite{Lutz:DISS,Athreya:ESDAICC}, have been
shown to have the useful characterizations
\begin{equation}\label{eq:1_1}
       \dim(S) = \liminf_{w\rightarrow S}\frac{\K(w)}{|w|\log |\Sigma|}
\end{equation}
and
\begin{equation}\label{eq:1_2}
       \Dim(S) = \limsup_{w\rightarrow S}\frac{\K(w)}{|w|\log |\Sigma|},
\end{equation}
where the logarithm is base-$2$ \cite{Mayordomo:KCCCHD,Athreya:ESDAICC}.  In these equations,
$\K(w)$ is the Kolmogorov complexity of the prefix $w$ of $S$, i.e., the
{\it length in bits of the shortest program} that prints the string
w.  (See \cite{LiVi97} for details.)  The numerators in
these equations are thus the {\it algorithmic information content} of
w, while the denominators are the ``naive'' information content of $w$,
also in bits.  We thus understand \eqref{eq:1_1} and \eqref{eq:1_2} to say that $\dim(S)$
and $\Dim(S)$ are the lower and upper {\it information densities} of the
sequence $S$.  These constructive dimensions and their analogs at other
levels of effectivity have been investigated extensively in recent
years \cite{EFD-bib}.

The constructive dimensions $\dim(S)$ and $\Dim(S)$ have recently been
generalized to incorporate a probability measure $\nu$ on the sequence
space $\Sigma^\infty$ as a parameter \cite{Lutz:DPSSF}.  Specifically, for each
such $\nu$ and each sequence $S \in \Sigma^\infty$, we now have the
constructive dimension $\dim^\nu(S)$ and the constructive strong
dimension $\Dim^\nu(S)$ of $S$ with respect to $\nu$.  (The first of these
is a constructive version of Billingsley dimension \cite{Bill60}.)  When
$\nu$ is the uniform probability measure on $\Sigma^\infty$, we have
$\dim^\nu(S) = \dim(S)$ and $\Dim^\nu(S) = \Dim(S)$.  A more interesting
example occurs when $\nu$ is the product measure generated by a
nonuniform probability measure $\beta$ on the alphabet $\Sigma$.  In this
case, $\dim^\nu(S)$ and $\Dim^\nu(S)$, which we write as $\dim^\beta(S)$ and
$\Dim^\beta(S)$, are again the lower and upper information densities of
S, but these densities are now measured with respect to unequal letter
costs.  Specifically, it was shown in \cite{Lutz:DPSSF} that
\begin{equation}\label{eq:1_3}
       \dim^\beta(S) = \liminf_{w\rightarrow S}\frac{\K(w)}{\I_\beta(w)}
\end{equation}
and
\begin{equation}\label{eq:1_4}
       \Dim^\beta(S) = \limsup_{w\rightarrow S}\frac{\K(w)}{\I_\beta(w)},
\end{equation}
where
\[ \I_\beta(w) = \sum_{i=0}^{|w|-1} \log\frac{1}{\beta(w[i])}\]
is the Shannon self-information of $w$ with respect to $\beta$.
These unequal letter costs $\log(1/\beta(a))$ for $a \in \Sigma$
can in fact be useful.  For example, the complete analysis of the
dimensions of individual points in self-similar fractals given by
\cite{Lutz:DPSSF} requires these constructive dimensions with a particular
choice of the probability measure $\beta$ on $\Sigma$.

In this paper we show how to use the constructive dimensions
$\dim^\beta(S)$ and $\Dim^\beta(S)$ in conjunction with randomness to
measure the degree to which two probability measures on $\Sigma$ are
similar.  To see why this might be possible, we note that the
inequalities
\[         0 \leq \dim^\beta(S) \leq \Dim^\beta(S) \leq 1\]
hold for all $\beta$ and $S$ and that the maximum values
\begin{equation}\label{eq:1_5}
             \dim^\beta(R) = \Dim^\beta(R) = 1
\end{equation}
are achieved if (but not only if) the sequence $R$ is random with respect to $\beta$.
It is thus reasonable to hope that, if $R$ is random with respect to
some other probability measure $\alpha$ on $\Sigma$, then $\dim^\beta(R)$ and
$\Dim^\beta(R)$ will take on values whose closeness to $1$ reflects the
degree to which $\alpha$ is similar to $\beta$.

This is indeed the case.  Our first main theorem says that the
{\it divergence formula}
\begin{equation}\label{eq:1_6}
 \dim^\beta(R) = \Dim^\beta(R) =\frac{\CH(\alpha)}{\CH(\alpha) + \D(\alpha||\beta)}
\end{equation}
holds whenever $\alpha$ and $\beta$ are computable, positive probability
measures on $\Sigma$ and $R \in \Sigma^\infty$ is random with respect to
$\alpha$.  In this formula, $\CH(\alpha)$ is the Shannon entropy of $\alpha$,
and $\D(\alpha||\beta)$ is the Kullback-Leibler divergence between $\alpha$
and $\beta$.  When $\alpha = \beta$, the Kullback-Leibler divergence
$\D(\alpha||\beta)$ is $0$, so \eqref{eq:1_6} coincides with \eqref{eq:1_5}.  When $\alpha$ and
$\beta$ are dissimilar, the Kullback-Leibler divergence $\D(\alpha||\beta)$
is large, so the right-hand side of \eqref{eq:1_6} is small.  Hence the
divergence formula tells us that, when $R$ is $\alpha$-random,
$\dim^\beta(R) = \Dim^\beta(R)$ is a quantity in $[0,1]$ whose closeness to
$1$ is an indicator of the similarity between $\alpha$ and $\beta$.

The proof of \eqref{eq:1_6}  serves as an outline of our other, more challenging
task, which is to prove that the divergence formula  \eqref{eq:1_6} also
holds for the much more effective {\it finite-state} $\beta$-{\it dimension}
$\dimfs^\beta(R)$ and {\it finite-state strong} $\beta$-{\it dimension}
$\Dimfs^\beta(R)$. (These dimensions are generalizations
of finite-state dimension and finite-state strong dimension, which were
introduced in \cite{Dai:FSD,Athreya:ESDAICC}, respectively.)

With this objective in mind, our second main theorem characterizes the
finite-state $\beta$-dimensions in terms of finite-state data compression.
Specifically, this theorem says that, in analogy with  \eqref{eq:1_3} and  \eqref{eq:1_4},
the identities
\begin{equation}\label{eq:1_7}
\dimfs^\beta(S) = \inf_C \liminf_{w\rightarrow S}\frac{|C(w)|}{\I_\beta(w)}
\end{equation}
and
\begin{equation}\label{eq:1_8}
       \dimfs^\beta(S) = \inf_C \limsup_{w\rightarrow S}\frac{|C(w)|}{\I_\beta(w)}
\end{equation}
hold for all infinite sequences $S$ over $\Sigma$.  The infima here are taken
over all information-lossless finite-state compressors (a model introduced
by Shannon \cite{Shannon48} and investigated extensively ever since) $C$ with output
alphabet ${0,1}$, and $|C(w)|$ denotes the number of bits that $C$ outputs when
processing the prefix $w$ of $S$.  The special cases of \eqref{eq:1_7} and \eqref{eq:1_8} in
which $\beta$ is the uniform probability measure on $\Sigma$, and hence
$\I_\beta(w) = |w| \log |\Sigma|$, were proven in \cite{Dai:FSD,Athreya:ESDAICC}.  In fact,
our proof uses these special cases as ``black boxes'' from which we derive
the more general \eqref{eq:1_7} and \eqref{eq:1_8}.

With \eqref{eq:1_7} and \eqref{eq:1_8} in hand, we prove our third main theorem.  This involves
the finite-state version of randomness, which was introduced by Borel \cite{Bore09}
long before finite-state automata were defined.  If $\alpha$ is a probability
measure on $\Sigma$, then a sequence $S \in \Sigma^\infty$ is $\alpha$-{\it normal}
in the sense of Borel if every finite string $w \in \Sigma^*$ appears with
asymptotic frequency $\alpha(w)$ in $S$, where we write
\[            \alpha(w) = \prod_{i=0}^{|w|-1} \alpha(w[i]).\]
Our third main theorem says that the {\it divergence formula}
\begin{equation}\label{eq:1_9}
\dimfs^\beta(R) = \Dimfs^\beta(R) =\frac{\CH(\alpha)}{\CH(\alpha) + \D(\alpha||\beta)}
\end{equation}
holds whenever $\alpha$ and $\beta$ are  positive probability
measures on $\Sigma$ and $R \in \Sigma^\infty$ is $\alpha$-normal.

\vspace{0.6cm}

\noindent{\bf Acknowledgments.}
I thank Xiaoyang Gu and Elvira Mayordomo for useful discussions.
\bibliographystyle{abbrv}

\begin{thebibliography}{}

\end{thebibliography}


\begin{thebibliography}{10}

\bibitem{Athreya:ESDAICC}
K.~B. Athreya, J.~M. Hitchcock, J.~H. Lutz, and E.~Mayordomo.
\newblock Effective strong dimension, algorithmic information, and
  computational complexity.
\newblock {\em SIAM Journal on Computing}, 37:671--705, 2007.

\bibitem{Bill60}
P.~Billingsley.
\newblock Hausdorff dimension in probability theory.
\newblock {\em Illinois Journal of Mathematics}, 4:187--209, 1960.

\bibitem{Bore09}
E.~Borel.
\newblock Sur les probabilit\'es d\'enombrables et leurs applications
  arithm\'etiques.
\newblock {\em Rend. Circ. Mat. Palermo}, 27:247--271, 1909.

\bibitem{CovTho06}
T.~M. Cover and J.~A. Thomas.
\newblock {\em Elements of Information Theory}.
\newblock John Wiley \& Sons, Inc., second edition, 2006.

\bibitem{Dai:FSD}
J.~J. Dai, J.~I. Lathrop, J.~H. Lutz, and E.~Mayordomo.
\newblock Finite-state dimension.
\newblock {\em Theoretical Computer Science}, 310:1--33, 2004.

\bibitem{Eggl49}
H.~Eggleston.
\newblock The fractional dimension of a set defined by decimal properties.
\newblock {\em Quarterly Journal of Mathematics}, Oxford Series 20:31--36,
  1949.

\bibitem{Haus19}
F.~Hausdorff.
\newblock Dimension und {\"a}usseres {M}ass.
\newblock {\em Mathematische Annalen}, 79:157--179, 1919.
\newblock English translation.

\bibitem{EFD-bib}
J.~M. Hitchcock.
\newblock Effective Fractal Dimension Bibliography,\\
\newblock \htmladdnormallink{http://www.cs.uwyo.edu/
  $\sim$jhitchco/bib/dim.shtml}{http://www.cs.uwyo.edu/\~jhitchco/bib/dim.shtm%
l} (current October, 2008).

\bibitem{LiVi97}
M.~Li and P.~M.~B. Vit\'{a}nyi.
\newblock {\em An Introduction to Kolmogorov Complexity and its Applications}.
\newblock Springer-Verlag, Berlin, 1997.
\newblock Second Edition.

\bibitem{Lutz:DCC}
J.~H. Lutz.
\newblock Dimension in complexity classes.
\newblock {\em SIAM Journal on Computing}, 32:1236--1259, 2003.

\bibitem{Lutz:DISS}
J.~H. Lutz.
\newblock The dimensions of individual strings and sequences.
\newblock {\em Information and Computation}, 187:49--79, 2003.

\bibitem{Lutz:DFRD}
J.~H. Lutz.
\newblock A divergence formula for randomness and dimension.
\newblock Technical Report cs.CC/0811.1825, Computing Research Repository,
  2008.

\bibitem{Lutz:DPSSF}
J.~H. Lutz and E.~Mayordomo.
\newblock Dimensions of points in self-similar fractals.
\newblock {\em SIAM Journal on Computing}, 38:1080--1112, 2008.

\bibitem{Martinlof66}
P.~Mar{tin-L{\"o}f}.
\newblock The definition of random sequences.
\newblock {\em Information and Control}, 9:602--619, 1966.

\bibitem{Mayordomo:KCCCHD}
E.~Mayordomo.
\newblock A {K}olmogorov complexity characterization of constructive
  {H}ausdorff dimension.
\newblock {\em Information Processing Letters}, 84(1):1--3, 2002.

\bibitem{Schn71a}
C.~P. Schnorr.
\newblock A unified approach to the definition of random sequences.
\newblock {\em Mathematical Systems Theory}, 5:246--258, 1971.

\bibitem{Schn77}
C.~P. Schnorr.
\newblock A survey of the theory of random sequences.
\newblock In R.~E. Butts and J.~Hintikka, editors, {\em Basic Problems in
  Methodology and Linguistics}, pages 193--210. D. Reidel, 1977.

\bibitem{Shannon48}
C.~E. Shannon.
\newblock A mathematical theory of communication.
\newblock {\em Bell System Technical Journal}, 27:379--423, 623--656, 1948.

\bibitem{Sull84}
D.~Sullivan.
\newblock Entropy, {Hausdorff} measures old and new, and limit sets of
  geometrically finite {Kleinian} groups.
\newblock {\em Acta Mathematica}, 153:259--277, 1984.

\bibitem{Tricot82}
C.~Tricot.
\newblock Two definitions of fractional dimension.
\newblock {\em Mathematical Proceedings of the Cambridge Philosophical
  Society}, 91:57--74, 1982.

\end{thebibliography}

\end{document}